# On the recent seismic activity at Kefalonia island (Greece): Manifestations of an Earth system in critical state.

## Y. Contoyiannis[1], S. M. Potirakis[2], J. Kopanas[3], G. Antonopoulos[3], G. Koulouras[4], K. Eftaxias[1], C. Nomicos[4].

1. Department of Physics, Section of Solid State Physics, University of Athens, Panepistimiopolis, GR-15784, Zografos, Athens, Greece, (Y. C.: yconto@yahoo.gr, K. E.: ceftax@phys.uoa.gr)
1. Department of Electronics Engineering, Technological Education Institute (TEI) of Piraeus, 250 Thivon & P. Ralli, GR-12244, Aigaleo, Athens, Greece, spoti@teipir.gr.
3. Department of Environmental Technology and Ecology, Technological Education Institute (TEI) of the Ionian Islands, Panagoulas road, GR-29100, Zante, Greece, (e- J. K.: jkopan@otenet.gr; e- G. A.: sv8rx@teiion.gr)
4. Department of Electronics Engineering, Technological Education Institute (TEI) of Athens, Ag. Spyridonos, GR-12210, Aigaleo, Athens, Greece, cnomicos@teiath.gr.

## Abstract

In this paper we show, in terms of fracture-induced electromagnetic emissions (EME) that the Earth system around the focal areas came to critical condition a few days before the occurrence of each one of the two recent earthquakes of Kefalonia (Cephalonia), Greece. Specifically, EME were recorded two days prior to the first earthquake [(38.22o N, 20.53oE), 26 January 2014, M=6.1] & six days prior to the second one [(38.26o N, 20.39oE), 03 February 2014, M=6.0]. Specifically, the MHz EME recorded by the remote telemetric stations on the island of Kefalonia and the neighboring island of Zante came simultaneously to critical condition in both cases. The analysis was performed by means of the method of critical fluctuations (MCF) revealing critical features.

## 1   Introduction

The possible connection of the electromagnetic (EM) activity that is observed prior to significant earthquakes (EQs) with the corresponding EQ preparation processes, often referred to as seismo-electromagnetics, has been intensively investigated during the last years. Several possible EQ precursors have been suggested in the literature [*Uyeda et al.*, 2009a; *Cicerone et al.*, 2009; *Hayakawa*, 2013a, 2013b]. The possible relation of the field observed fracture-induced electromagnetic emissions (EME) in the frequency bands of MHz and kHz has been examined in a series of publications [e.g., *Eftaxias et al.*, 2001, 2004, 2008; *Kapiris et al.*, 2004; *Karamanos et al.,* 2006; *Papadimitriou et al.*, 2008; *Contoyiannis et al.,* 2005, 2013; *Eftaxias and Potirakis*, 2013a; *Potirakis et al.*, 2011, 2012a, 2012b, 2012c, 2013; *Minadakis et al.*, 2012a, 2012b], while a three stage model for the preparation of an EQ by means of its observable EM activity has been recently put forward [*Eftaxias and Potirakis*, 2013b, and references therein].

In this letter, we report the recording, with a sampling rate of 1 sample/s, of two pairs of MHz EM signals: one pair prior to each one of the recent significant EQs occurred in south-west Greece. On 26 January 2014 (13:55:43 UT) an $M = 6.1$ EQ occurred on the Island of Kefalonia (Cephalonia), while 2 days before (on 24 January 2014) two remote telemetric stations of our remote observation stations network, the station of Kefalonia (located on the same island) and the station of Zante (located on a neighboring island of the same island complex) recorded simultaneously the first pair of aforementioned signals. The same picture was repeated for the second significant EQ, $M = 6.0$, that occurred on the same island on 3 February 2014 (03:08:46 UT). Specifically, both the Kefalonia and the Zante station recorded simultaneously the second pair of aforementioned signals six days prior to the specific EQ (on 28 January 2014). Note that, it has been repeatedly made clear that the pre-EQ EME signals have been recorded only prior to strong shallow EQs that happened in land (or near coast-line); this also, in combination to the recently proposed fractal geo-antenna model [*Eftaxias et al.*, 2004], explains why they manage to be transmitted to the air. This model gives a good reason for the increased possibility of detection of such EM radiation, since it has as a consequence an increased radiated power compared to the power that would be radiated if a dipole antenna model was considered.

The analysis of the specific EM time-series, using the method of critical fluctuations (MCF) [*Contoyiannis and Diakonos*, 2000; *Contoyiannis et al.*, 2002, 2013], reveals critical features, implying that the possibly related underlying geophysical process is at critical state. The presence of the "critical point" during which any two active parts of the system are highly correlated even at arbitrarily long distances, in other words when "everything depends on everything else", is consistent with the view that the EQ preparation process during the period that the MHz EME are emitted is a spatially extensive process. It is noted that, according to the aforementioned three stage model [*Eftaxias and Potirakis*, 2013b, and references therein], the pre-seismic MHz EM emission is considered to be originate during the fracture of the part of the Earth's crust that is characterized by highly heterogeneity. During this phase the fracture is non-directional and spans over a large area that surrounds the family of large high-strength entities distributed along the fault sustaining the system. Note that for an EQ of magnitude ~6 the corresponding fracture process extends to a radius of ~120km [*Bowman et al.*, 1998].

## 2   Data analysis method

The analysis of the recorded data was performed using the method of critical fluctuations (MCF) [*Contoyiannis and Diakonos*, 2000; *Contoyiannis et al.*, 2002, 2013]. Detailed descriptions of all the involved calculations can be found elsewhere [*Contoyiannis et al.*, 2013] and therefore are omitted here for the sake of brevity and focus on the findings. However, a general description of the employed method follows.

MCF was proposed for the analysis of critical fluctuations in the observables of systems that undergo a continuous phase transition [*Contoyiannis and Diakonos*, 2000; *Contoyiannis et al.*, 2002]. It is based on the finding that the fluctuations of the order parameter, that characterizes successive configurations of critical systems at equilibrium, obey a dynamical law of intermittency of an 1D nonlinear map form. The MCF is applied to stationary time windows of statistically adequate length, for which the distribution of the of waiting times $l$ (laminar lengths) of fluctuations in a properly defined laminar region is fitted by a function $f(l) \propto l^{-p_2} e^{-p_3 l}$. The criteria for criticality are $p_2 > 1$ and $p_3 \approx 0$ [*Contoyiannis and Diakonos*, 2000; *Contoyiannis et al.*, 2002]. In that case the system is characterized by intermittent dynamics, since the distribution follows power-law decay [*Schuster*, 1998]. On the other hand, in the case of a system governed by noncritical dynamics the corresponding distribution follows an exponential decay, rather than a power-law one [*Contoyiannis et al.*, 2004b]. The MCF has been applied to a variety of dynamical systems, including thermal (e.g., 3D Ising) [*Contoyiannis et al.*, 2002], geophysical [*Contoyiannis et al.*, 2004a; *Contoyiannis and Eftaxias* 2008; *Contoyiannis et al.*, 2010] and biological systems (electro-cardiac signals) [*Contoyiannis et al.*, 2004b; *Contoyiannis et al.*, 2013].

## 3    Analysis results

Part of the MHz recordings of the Kefalonia station associated with the $M = 6.1$ EQ is shown in Fig. 1a. This was recorded in Julian day 24, that is ~2 days before the occurrence of the first Kefalonia EQ. This stationary time-series excerpt, having a total length of 2.8h (10000 samples) starting at 24 Jan. 2014 (12:46:40 UT), was analyzed by the MCF method and was identified to be a "critical window" (CW). CWs are time intervals of the MHz EME signals presenting features analogous to the critical point of a second order phase transition [*Contoyiannis et al.*, 2005].

The main steps of the MCF analysis [*Contoyiannis et al.*, 2013] on the specific time-series are shown in Fig. 1b- Fig. 1d. First, a distribution of the amplitude values of the analyzed signal was obtained from which, using the method of turning points [*Pingel et al.*, 1999], a fixed-point, that is the start of laminar regions, $\phi_o$ of about 700mV was determined. Fig. 1c portrays the obtained laminar distribution for the end point $\phi_l = 655mV$, that is the distribution of waiting times, referred to as laminar lengths $l$, between the fixed-point $\phi_o$ and the end point $\phi_l$, as well as the fitted function $f(l) \propto l^{-p_2} e^{-p_3 l}$ with the corresponding exponents $p_2 = 1.35$, $p_3 = 0.000$ with $R^2 = 0.999$. Finally, Fig. 1d shows the obtained plot of the $p_2$, $p_3$ exponents vs. $\phi_l$. From Fig. 1d it is apparent that the criticality conditions, $p_2 > 1$ and $p_3 \approx 0$, are satisfied for a wide range of end points $\phi_l$, revealing the power-law decay feature of the time-series that proves that the system is characterized by intermittent dynamics; in other words, the MHz time-series excerpt of Fig. 1a is indeed a CW.

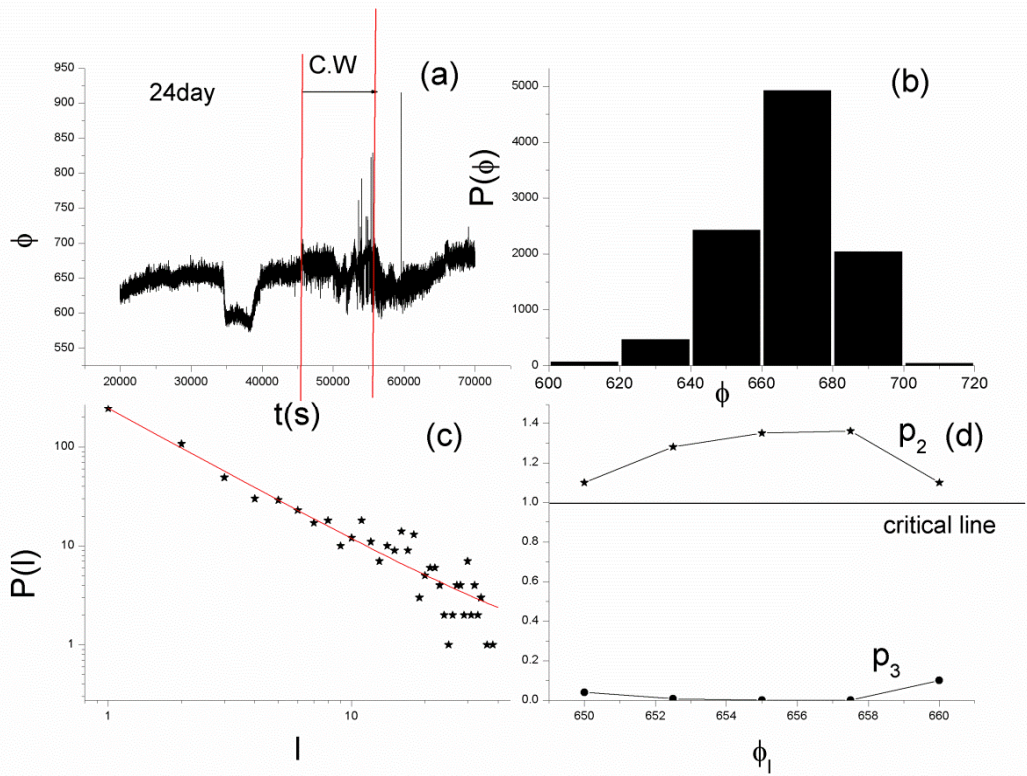

Fig. 1. (a) The 10000 samples long critical window of the MHz EME that was recorded before the Kefalonia $M = 6.1$ EQ at the Kefalonia station. (b) Amplitude distribution of the signal of Fig. 1a. (c) Laminar distribution for the end point $\phi_l = 655mV$ , as a representative example of the involved fitting. The solid line corresponds to the fitted function (cf. to text in Sec. 2) with the values of the corresponding exponents $p_2$, $p_3$ also noted. (d) The obtained exponents $p_2$, $p_3$ vs. different values of the end of laminar region $\phi_l$. The horizontal dashed line indicates the critical limit ( $p_2 = 1$ ).

Part of the MHz recordings of the Zante station associate with the $M = 6.1$ EQ is shown in Fig. 2a. This was also recorded in Julian day 24, that is ~2 days before the occurrence of Kefalonia EQ. This stationary time-series excerpt, having a total length of 2.8h (10000 samples) starting at 24 Jan. 2014 (12:46:40 UT), was also analyzed by the MCF method and was identified to be a "critical window" (CW).

The main steps of the MCF analysis [*Contoyiannis et al.*, 2013] on the specific time-series are shown in Fig. 2b- Fig. 2d. First, a distribution of the amplitude values of the analyzed signal was obtained from which, using the method of turning points [*Pingel et al.*, 1999], a fixed-point, that is the start of laminar regions, $\phi_o$ of about 600mV was determined. Fig. 2c portrays the obtained laminar distribution for the end point $\phi_l = 665mV$ , that is the distribution of waiting times, referred to as laminar lengths $l$, between the fixed-point $\phi_o$ and the end point $\phi_l$, as well as the fitted function $f(l) \propto l^{-p_2} e^{-p_3 l}$ with the corresponding exponents $p_2 = 1.49$, $p_3 = 0.000$ with $R^2 = 0.999$. Finally, Fig. 2d shows the obtained plot of the $p_2$, $p_3$ exponents vs. $\phi_l$. From Fig. 2d it is apparent that the criticality conditions, $p_2 > 1$ and $p_3 \approx 0$, are satisfied for a wide range of end points $\phi_l$, revealing the power-law decay feature of the time-series that proves that the system is characterized by intermittent dynamics; in other words, the MHz time-series excerpt of Fig. 2a is indeed a CW.

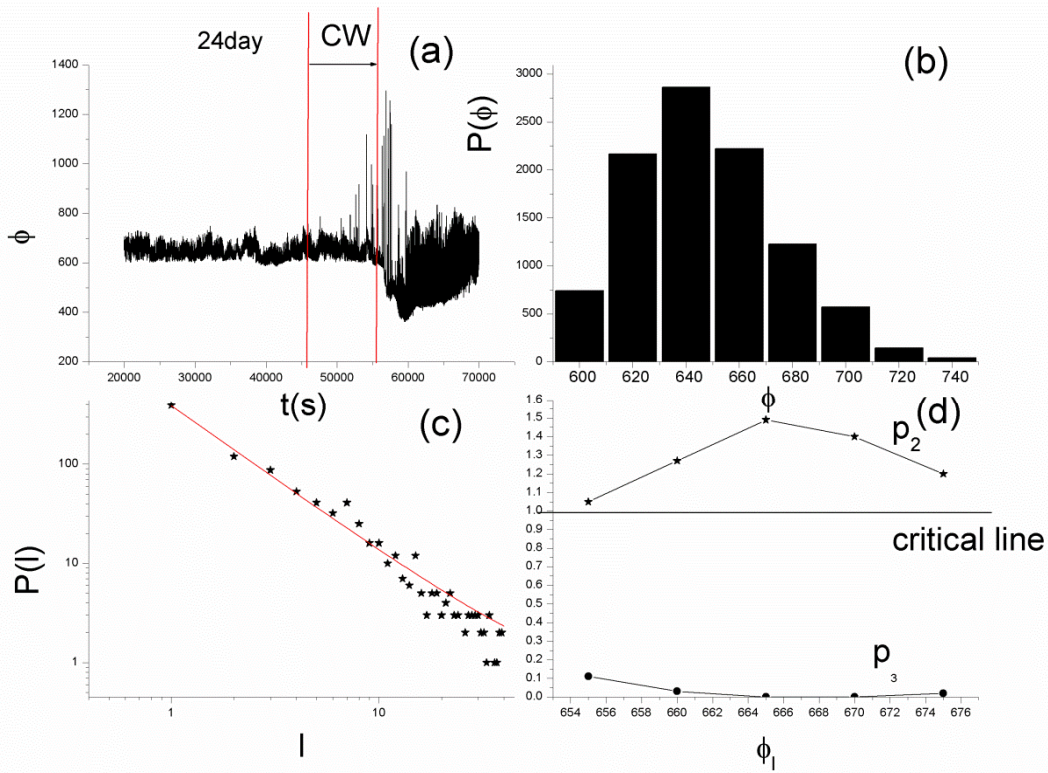

Fig. 2. (a) The 10000 samples long critical window of the MHz EME that was recorded prior to the Kefalonia $M = 6.1$ EQ at the Zante station. (b) Amplitude distribution of the signal of Fig. 2a. (c) Laminar distribution for the end point $\phi_l = 665 mV$, as a representative example of the involved fitting. The solid line corresponds to the fitted function (cf. to text in Sec. 2) with the values of the corresponding exponents $p_2$, $p_3$ also noted. (d) The obtained exponents $p_2$, $p_3$ vs. different values of the end of laminar region $\phi_l$. The horizontal dashed line indicates the critical limit ($p_2 = 1$).

After the $M = 6.1$ EQ, ~ a week later, the second, $M = 6.0$, EQ occurred on the same island with a focal area a few km further than the first one. Six days earlier, the Kefalonia station recorded MHz EME while the Zante station also recorded at the same time.

Specifically, a stationary time-series excerpt, having a total length of 3.3h (12000 samples) starting at 28 Jan. 2014 (05:33:20 UT), from each station was analyzed by the MCF method and both of them were identified to be CWs. Figs 3 & 4 show the results of the correxponding analyses.

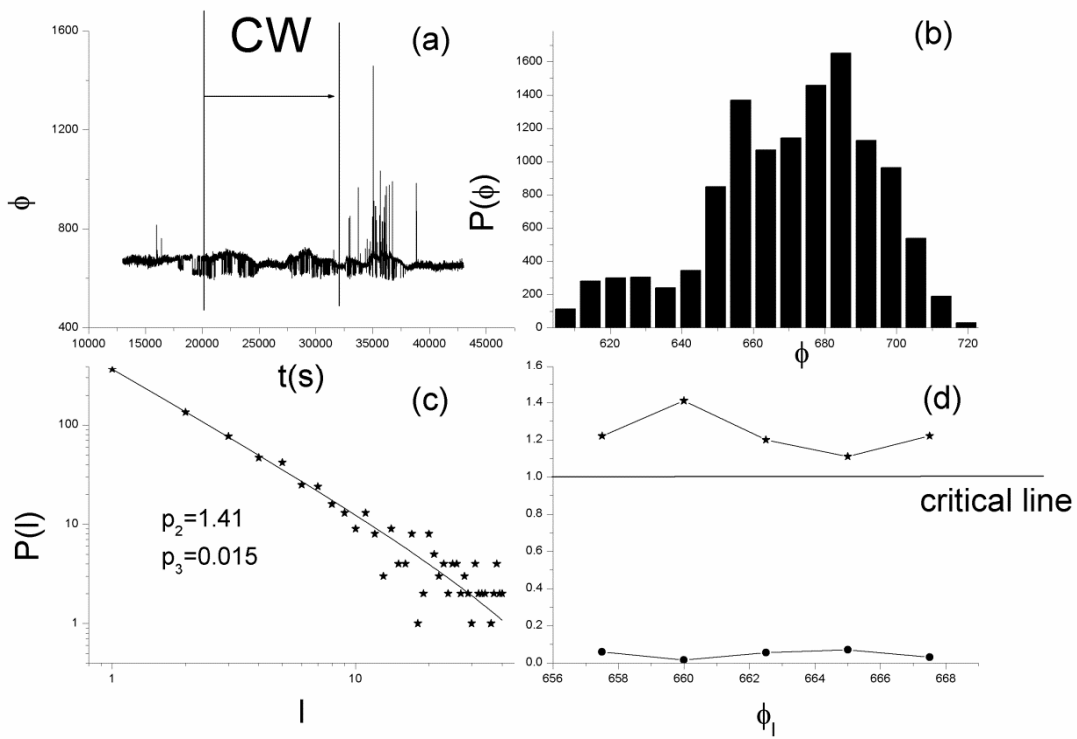

Fig. 3. (a) The 12000 samples long critical window of the MHz EME that was recorded before the Kefalonia $M = 6.0$ EQ at the Kefalonia station. (b) Amplitude distribution of the signal of Fig. 3a. (c) Laminar distribution: a representative example of the involved fitting. The solid line corresponds to the fitted function (cf. to text in Sec. 2) with the values of the corresponding exponents $p_2$, $p_3$ also noted. (d) The obtained exponents $p_2$, $p_3$ vs. different values of the end of laminar region $\phi_l$. The horizontal dashed line indicates the critical limit ($p_2 = 1$).

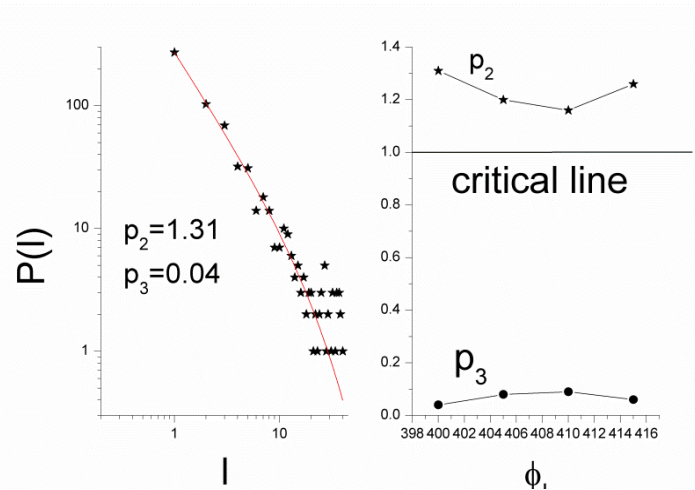

Fig. 4. Left: The laminar distribution, as a representative example of the involved fitting, for the MHz signal recorded at the Zante station prior to the Kefalonia $M = 6.0$ EQ. The solid line corresponds to the fitted function (cf. to text in Sec. 2) with the values of the corresponding exponents $p_2$, $p_3$ also noted. Right: The obtained exponents $p_2$, $p_3$ vs. different values of the end of laminar region $\phi_l$. The horizontal dashed line indicates the critical limit ($p_2 = 1$).

In summary, we conclude that both stations recorded MHz signals that simultaneously presented critical state features two days before the first main event and six days before the second main event.

## 4  Discussion - Conclusions

Based on the method of critical fluctuations, we have shown that the fracture-induced MHz EME recorded by two stations of our remote observation stations network prior to the two recent significant EQs of Kefalonia present criticality characteristics, implying that they emerge from a system in critical state.

There are two key points that render these observations unique in the up to now research on the preseismic EME:

(i) The Kefalonia station is known for being insensitive to EQ preparation processes happening outside of the wider area of Kefalonia island, as well as to EQ preparation processes leading to low magnitude EQs within the area of Kefalonia island. Note that the only signal that has been previously recorded refers to the M=6 EQ that occurred on the specific island in 2007 [*Contoyiannis et al.*, 2010].

(ii) MHz EME presenting critical characteristics were simultaneously recorded in two different stations very close to the focal area, while no other station of our network (cf. Fig. 3) has recorded such signals prior to the specific EQ. This feature, combined with the abovementioned sensitivity of the Kefalonia station only to significant EQs occurring on the specific island, may be considered as an indication of the location of the impending EQ.

We note that, seismicity and prefracture EM emissions should be two sides of the same coin concerning the earthquake generation process. Therefore, the corresponding foreshock seismic activity, as another manifestation of the same complex system, should also be at critical state as well. We have shown that this really happens [Potirakis et al., 2013].

EME, as a phenomenon rooted in the damage process, should be an indicator of memory effects. Laboratory studies verify that: during cyclic loading, the level of EME increases significantly when the stress exceeds the maximum previously reached stress level (Kaizer effect). The existence of Kaizer effect predicts the EM silence during the aftershock period. Thus, the appearance of the second EM anomaly reveals that the corresponding preparation of fracture process has been organized in a new barrier.

We finally note that the multidisciplinary analysis of the kHz EME recorded our stations is currently in progress. Indications in terms of traceability of the observed kHz EM emissions support the hypothesis that the detected EM anomalies are associated with the seismic activity under study.

As it has been repeatedly pointed out in our works, our view is that such observations and the associated analyses offer valuable information for the comprehension of the Earth system processes that take place prior to the occurrence of a significant EQ. As it is known a large number of other precursory phenomena are also observed, both by ground and satellite stations, prior to significant EQs. Only a combined evaluation of our observations with other well documented precursory phenomena could possibly render our observations useful for a reliable short-term forecast solution. In the cases under study this requirement was not fulfilled.

Fig. 3. The 11 remote sensing stations completing the telemetric network for the recording of electromagnetic variations in the MHz and kHz bands in Greece.

## Acknowledgements

Research co-funded by the EU (European Social Fund) and national funds, action "Archimedes III—Funding of research groups in T.E.I.", under the Operational Programme "Education and Lifelong Learning 2007-2013".